# Antiparticle in Light of Einstein-Podolsky-Rosen Paradox and Klein Paradox [*]

Guang-jiong Ni[†]
*Department of Physics, Fudan University, Shanghai, 200433, P. R. China*

Hong Guan[‡]
*Department of Physics, Zhongshan University, Guangzhou, 510275, P. R. China*

Weimin Zhou[§] and Jun Yan[**]
*Department of Physics, New York University, 4 Washington Place, New York, NY, 10003*

The original version of Einstein-Podolsky-Rosen (EPR) paradox and the Klein paradox of Klein-Gordon (KG) equation are discussed to show the necessity of existence of antiparticle with its wavefunction being fixed unambiguously. No concept of "hole" is needed.
PACS: 03.65-w, 03.65.Bz, 03.65.Pm

Every particle has its antiparticle. However, in most textbooks on quantum mechanics (QM), there is nearly no clear explanation for the wavefunction (WF) of antiparticle. Some times (and often implicitly), it was said that the WF of antiparticle is the same as that of particle, one has to distinguish the WF of positron from that of electron by adding some words. Alternatively, one often said that in the vacuum all the negative states ("sea") of electron are filled. Once a "hole" is created in the "sea", it would correspond to a positron.

Evidently, the concept of "sea" and "hole" can not work in the case of Klein-Gordon (KG) equation, where it is hopeless to fill the doubly infinite states of negative energy.

In this letter, through the investigation on the original version of Einstein- Podolsky-Rosen (EPR) paradox [1] and the Klein paradox [2] of KG equation, we will show the essential necessity of introducing the antiparticle state to QM and the form of its WF being fixed unambiguously. No "hole" concept is needed.

The basic idea was explained in references [3]-[7]. A particle is always not pure. It always contains two contradicted fields, $\varphi(\vec{x},t)$ and $\chi(\vec{x},t)$. For instance, the KG equation

$$[\hbar\frac{\partial}{\partial t} - V(x)]^2\psi = -c^2\hbar^2\nabla^2\psi + m^2c^4\psi \qquad (1)$$

can be recast into two coupled Schrödinger equation of $\varphi$ and $\chi$ [8]:

$$\begin{cases} (\hbar\frac{\partial}{\partial t} - V)\varphi &= m_c^2\varphi - \frac{\hbar^2}{2m}\nabla^2(\varphi+\chi) \\ (\hbar\frac{\partial}{\partial t} - V)\chi &= -mc^2\chi + \frac{\hbar^2}{2m}\nabla^2(\chi+\varphi) \end{cases} \qquad (2)$$

with

$$\begin{cases} \varphi &= \frac{1}{2}[(1-\frac{V}{mc^2})\psi + i\frac{\hbar}{mc^2}\dot\psi] \\ \chi &= \frac{1}{2}[(1+\frac{V}{mc^2})\psi - i\frac{\hbar}{mc^2}\dot\psi] \end{cases} \qquad (3)$$

The Eq. (2) is invariant under the (newly defined) space-time inversion ($\vec{x} \to -\vec{x}, t \to -t$) and transformation:

$$\varphi(-\vec{x},-t) \to \chi(\vec{x},t), \qquad \chi(-\vec{x},-t) \to \varphi(\vec{x},t) \qquad (4)$$

$$V(-\vec{x},-t) \to -V(\vec{x},t) \qquad (5)$$

However, for a concrete solution of WF for a particle, e.g., the plane wave:

$$\psi \sim \exp[\frac{i}{\hbar}(\vec{p}\cdot\vec{x} - Et)] \qquad (6)$$

---

[*]Supported in part by the National Science Foundation: 19910210076-A05
[†]E-mail: gjni@fudan.ac.cn
[‡]E-mail: stdp@zsulink.zsu.edu.cn
[§]E-mail: wz214@is6.nyu.edu
[**]E- mail: jy272@scires.nyu.edu

the symmetry (4) is not explicit but hidden. This is because $|\varphi| > |\chi|$ with $\chi$ being the "slave" who has to obey the "master" $\varphi$ and follow the space-time evolution of $\varphi$ as shown at Eq. (6).

Let us perform a space-time inversion on the particle state, then its WF (6) transforms into

$$\psi_c \sim \exp[-\frac{i}{\hbar}(\vec{p}\cdot\vec{x} - Et)] \tag{7}$$

with $\chi_c(\vec{x},t) = \varphi(-\vec{x},-t)$ becoming the "master" whereas $\varphi_c(\vec{x},t) = \chi(-\vec{x},-t)$ reduced into "slave" since now $|\chi_c| > |\varphi_c|$. The Eq. (7) describes the WF of an antiparticle with the momentum $\vec{p}$ and energy $E > 0$ precisely the same as that in the particle state (6).

What we said above can be generalized into a postulate that "the space-time inversion ($\vec{x} \to -\vec{x}, t \to -t$) is equivalent to the transformation between the particle and its antiparticle". In our opinion, this is an inevitable outcome of the long study on the C, P, T problems since the historic discovery of parity (P) violation [9], [10]. the so-called "time reversal" symmetry (T) is violated whereas CPT theorem remains valid as directly verified by recent experiments. The transformation of a particle state $|a\rangle$ to its antiparticle state $|\bar{a}\rangle$ is not the charge conjugate transform C ($e \to -e$) but the CPT combined transform [12]:

$$|\bar{a}\rangle = CPT|a\rangle \tag{8}$$

The complex conjugation ($\psi \to \psi^*$) contained in C transform just cancels that in T transform and what Eq. (8) means is nothing but the transformation from Eq. (6) to (7).

We are now in a position to restudy the EPR paradox [1], which actually comprises two aspects. One of them involves the puzzle of entangled state, i.e., the nonlocally correlated two-particle quantum state over long distance and was verified again and again by experiments, especially in recent year [13], [14]. however, the another puzzle in EPR paper was overlooked by the majority of physicists but reemphasized in Ref. [15] as follows:

Consider two spinless particles in one dimensional space with positions $x_i$ ($i = 1, 2$) and momentum operators $\hat{p}_i = -i\hbar\frac{\partial}{\partial x_i}$. The $(x_1 - x_2)$ and $(\hat{p}_1 + \hat{p}_2)$ will commute:

$$[x_1 - x_2, \hat{p}_1 + \hat{p}_2] = 0 \tag{9}$$

which means that they have a common eigenstate with eigenvalues

$$x_1 - x_2 = a = const \tag{10}$$

$$p_1 + p_2 = 0, \quad p_2 = -p_1 \tag{11}$$

How strange the state is!? Two particles are moving in opposite momentum directions while keeping their distance unchanged. Incredible! As stressed in Ref. [15], "No one can figure out how to realize it". Now we propose the following answer.

If the WF of particle 1 is written as Eq. (6)

$$\psi(x_1,t) = \exp\{\frac{i}{\hbar}(p_1 x_1 - E_1 t)\} \tag{12}$$

with momentum $p_1 > 0$ and energy $E_1 > 0$, then the particle 2 must be an antiparticle:

$$\psi_c(x_2,t) = \exp\{\frac{i}{\hbar}(p_2 x_2 - E_2 t)\} = \exp\{-\frac{i}{\hbar}(p_1 x_2 - E_1 t)\} \tag{13}$$

with $p_2 = -p_1 < 0$, $E_2 = -E_1 < 0$. But as shown at Eq. (7), we should view the WF of "negative energy state" of particle directly as the WF of its antiparticle with the corresponding operators:

$$\hat{p}_c = \hbar\frac{\partial}{\partial x}, \qquad \hat{E}_c = -\hbar\frac{\partial}{\partial t} \tag{14}$$

which are the counterparts of those for the particle:

$$\hat{p} = -\hbar\frac{\partial}{\partial x}, \qquad \hat{E} = \hbar\frac{\partial}{\partial t} \tag{15}$$

So the observed momentum and energy of antiparticle in state (13) are $p_1$ and $E_1$ respectively, precisely the same as that of the particle state (12). Now everything is reasonable.

If instead of (9), we consider the commutation relation:

$$[x_1 + x_2, \hat{p}_1 - \hat{p}_2] = 0 \tag{16}$$

Then the correct answer turns out to be a particle and its antiparticle (C) moving in opposite directions with momentum $p_1$ versus $p_2^{(c)} = -p_1$ and position $x_1$ versus $x_2 = -x_1$. Such kind of experiments have been performed for many times, say, in the $e^+e^-$ pair creation by a high energy photon in the vicinity of heavy nucleus. Especially, a recent experiment reveals an entangled state of $K^0 - \bar{K}^0$ [16] system and provides a beautiful realization of relation (16).

Next, we turn to the KG Eq. (1) in one dimensional space with the potential barrier:

$$V(x) = \begin{cases} 0 & x < 0 \\ V_0 & x > 0 \end{cases} \tag{17}$$

The incident wave, reflected wave and transmitted wave are

$$\psi_i = a\exp[\frac{i}{\hbar}(px - Et)] \qquad (x < 0) \tag{18}$$

$$\psi_r = b\exp[\frac{i}{\hbar}(-px - Et)] \qquad (x < 0) \tag{19}$$
$$\psi_t = b'\exp[\frac{i}{\hbar}(p'x - Et)] \qquad (x > 0) \tag{20}$$

respectively with $p^2 = E^2/c^2 - m^2c^2$ and $p'^2 = (E-V_0)^2/c^2 - m^2c^2$. The continuation condition leads to

$$\frac{b}{a} = \frac{p-p'}{p+p'} \qquad \frac{b'}{a} = \frac{2p}{p+p'} \tag{21}$$

The Klein paradox emerges when $V_0 > E + mc^2$. Since now $p' = \pm\sqrt{(V_0-E)^2/c^2 - m^2c^2}$ remains real, the transmitted wave is oscillating while the reflectivity of incident wave reads

$$R = \left|\frac{b}{a}\right|^2 = \frac{|p-p'|^2}{|p+p'|^2} = \begin{cases} < 1 & \text{if } p' > 0 \\ > 1 & \text{if } p' < 0 \end{cases} \tag{22}$$

While the choice of $p' > 0$ is obviously unreasonable, the choice of $p' < 0$ seems quite attractive. Then we have to understand why $p' < 0$ and what happens in this situation?

For this purpose, we look back at Eqs. (1)-(6) and the accompanying continuity equation

$$\frac{\partial \rho}{\partial t} + \nabla \cdot \vec{j} = 0 \tag{23}$$

$$\rho = \frac{\hbar}{2mc^2}(\psi^*\dot{\psi} - \psi\dot{\psi}^*) - \frac{V}{mc^2}\psi^*\psi = \varphi^*\varphi - \chi^*\chi \tag{24}$$

$$\begin{aligned}\vec{j} &= \frac{\hbar}{2m}(\psi\nabla\psi^* - \psi^*\nabla\psi) \\ &= \frac{\hbar}{2m}[(\varphi\nabla\varphi^* - \varphi^*\nabla\varphi) + (\chi\nabla\chi^* - \chi^*\nabla\chi) \\ &\quad + (\varphi\nabla\chi^* - \chi^*\nabla\varphi) + (\chi\nabla\varphi^* - \varphi^*\nabla\chi)]\end{aligned} \tag{25}$$

Combining them with Eqs. (17)-(22), we find

$$\rho_i = |\varphi_i|^2 - |\chi_i|^2 = \frac{E}{mc^2}|a|^2 > 0, \qquad j_i = \frac{p}{m}|a|^2 > 0 \tag{26}$$

$$\rho_r = \frac{E}{mc^2}|b|^2 > 0, \qquad j_i = -\frac{p}{m}|b|^2 < 0 \tag{27}$$

$$\rho_t = |\varphi_t|^2 - |\chi_t|^2 = \frac{E-V_0}{mc^2}|b'|^2 < 0, \qquad j_t = \frac{p'}{m}|b'|^2 \tag{28}$$

So we should demand $p' < 0$ to get $j_t < 0$ in conformity with $\rho_t < 0$ and to meet the requirement of Eq. (23)($j_i + j_r = j_t$) with $|j_r| > j_i$ ($|b| > |a|$) and reflectivity $R > 1$.

The reason is clear. For an observer located at $x > 0$, the energy of particle in the transmitted wave should be measured with respect to the local potential $V_0$. In other words, the particle has energy $E' = E - V_0$ locally. Hence the wave function should be redefined as:

$$\psi_t \to \psi_t^{(c)} = b' \exp[\frac{i}{\hbar}(p'x - E't)] = b' \exp[-\frac{i}{\hbar}(|p'|x - |E'|t)] \qquad (x > 0) \qquad (29)$$

According to Eq. (14), the energy and momentum of this antiparticle are $|E'| > 0$ and $|p'| > 0$ respectively. It moves to the right though $p' < 0$ and $j_t < 0$. Therefore, the KG equation is self-consistent even at one-particle level to explain qualitatively the phenomenon of $\pi^+\pi^-$ pair creation at strong field barrier bombarded by incident $\pi^-$ beam.

In summary:

A. The original version of EPR paradox and Klein paradox in KG equation provide two convincing arguments for the necessary existence of antiparticle with its WF given by Eq. (7) and corresponding operators (14). The historic mission of "hole" concept is coming to an end. Historically, similar point of view had been considered by Schwinger [17]), Konopinski and Mahmand [18], Stueckelberg [19] and Feynman [20].

B. The CPT theorem already exhibits itself as a basic postulate. The transformation of a particle to its antiparticle($\mathcal{C}$) is not something which can be defined independently but a direct consequence of (newly defined) space-time inversion $\mathcal{PT}$ ($\vec{x} \to -\vec{x}, t \to -t$).

$$\mathcal{PT} = \mathcal{C} \qquad (30)$$

As a comparison, though the Newton's equation $F = ma$ could be derived from Lagrange variational principle or Hamilton Principle, it is a law rather than a theorem. The definition of inertial mass, $m$, is given by the equation $m = F/a$ which is verified by experiments.

C. The basic symmetry should be pushed forward to Eq. (4), forming a starting point to construct the theory of special relativity (SR), the relativistic QM and the quantum field theory (QFT). Three examples are as follows:

a) In some sense, the time reading of the "clock" for $\varphi$ field is "clockwise" whereas that for $\chi$ field is "anticlockwise" [ as shown by Eq.(6 versus 7)] essentially. Hence, in a particle state with $|\varphi| > |\chi|$, though the time reading remains "clockwise", it runs slower and slower due to the enhancement of hiding $\chi$ field accompanying with the increase of particle velocity.

b) As shown in [7], there is an upper bound for the velocity of KG or Dirac particle in free motion: $v_{max} \to c$, also a lower bound for its total energy $E$ in an external field: $E_{min} \geq 0$. Both these bounds are fixed by the upper limit for the ratio of hiding ingredient of $\chi$ to that of $\varphi$: $R = \int |\chi|^2 d\vec{x} / \int |\varphi|^2 d\vec{x} \to 1$. If ignoring all the $\chi$ field terms in the KG equation, one goes back to the Schrödinger equation where neither upper bound for velocity nor lower bound for energy exists.

c) In QFT, one starts from the field operator for KG field defined as

$$\hat{\psi}(\vec{x}, t) = \sum_{\vec{p}} \frac{1}{\sqrt{2V\omega_p}} \left[ \hat{a}_{\vec{p}} e^{\frac{i}{\hbar}(\vec{p}\cdot\vec{x} - Et)} + \hat{b}_{\vec{p}}^{\dagger} e^{-\frac{i}{\hbar}(\vec{p}\cdot\vec{x} - Et)} \right] \qquad (31)$$

The correctness of $\hat{\psi}$ is ensured by its invariance under space-time inversion, $\hat{\psi}(-\vec{x}, -t) = \hat{\psi}(\vec{x}, t)$, which contains the transformation

$$\hat{a}_{\vec{p}} \rightleftharpoons \hat{b}_{\vec{p}}^{\dagger} \qquad (\vec{x} \to -\vec{x}, t \to -t) \qquad (32)$$

as a intuitive complement of Eq. (30).

D. What we are discussing is the essential conformity between QM and SR. In some sense the subtle relationship between them lies in the essential equivalence of "$+i$" and "$-i$". Indeed, at the very elementary level of physics, " that not forbidden is allowed" (M. Gell-Mann). It is the time for the revival of "Ether".